# Defining structural gradient hardening through Type II back stress for heterostructured materials

En Ma[1] and Ting Zhu[2]

[1]Center for Alloy Innovation and Design (CAID), Xi'an Jiaotong University, Xi'an, 710049, China

[2]Woodruff School of Mechanical Engineering, Georgia Institute of Technology, Atlanta, GA 30332, USA

**Email:** maen@xjtu.edu.cn (E.M.); ting.zhu@me.gatech.edu (T.Z.)

## Abstract

The recently proposed term "heterostructured (HS) materials" serves as an umbrella classification encompassing a wide range of materials that hold great promise for enhanced mechanical properties. Most HS materials exhibit back-stress strengthening, as is typical for all plastically non-homogeneous materials. To better embody the distinctiveness of materials crafted via innovative heterostructuring, here we introduce the concept of "structural gradient hardening" (SGH), which captures an essential feature of HS materials and complements traditional strengthening mechanisms. SGH refers to the extra strengthening that arises from a characteristic gradient structure introduced by heterostructuring, beyond what is predicted by the rule of mixtures. This distinction is useful, as the overall back stress can in fact be partitioned into Type I and Type II components, with the latter specifically quantifying the extra hardening originating from the structural and strain gradients established by heterostructuring, as articulated in this Viewpoint article.

**Heterostructured materials**

Recent years have witnessed growing research on "heterostructured (HS) materials" [1], a topic that has become the focus of recurring symposia at many materials science and engineering conferences. An increasing number of publications now adopt this terminology [1-3], most of which investigate the plastic deformation behavior of HS materials. These studies often attribute the simultaneously high strength and ductility observed in HS materials to heterogeneous deformation induced (HDI) strengthening [1]. The latter is typically quantified by the experimentally measured sample-level back stress ($\sigma_b$), which reflects directional, long-range internal stresses generated by the accumulation of geometrically necessary dislocations (GNDs) at boundaries between zones with inhomogeneous deformation or at strengthening obstacles. Placed in historical context, the concepts of back stresses and GNDs arising from heterogeneity can be traced back to the 1960s and 1970s [4-7]. Fig. 1 schematically illustrates this well-established strengthening mechanism [8].

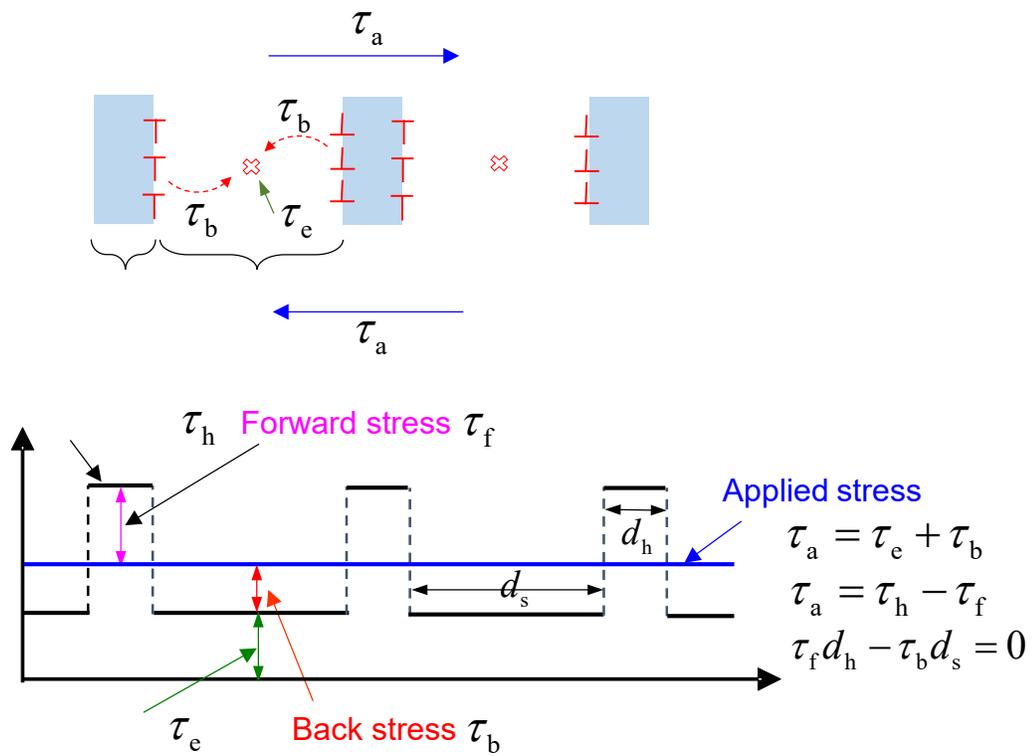

**Figure 1**. Schematic illustration of HDI strengthening through the buildup of internal shear stresses, including back stress $\tau_b$ and forward stress $\tau_f$, in a composite of hard and soft domains under applied shear stress $\tau_a$, adapted from [8]. (a) GNDs, represented by ⊥, accumulate at the interfaces between hard and soft domains to accommodate deformation incompatibility. These GNDs generate back stresses $\tau_b$, which hinder the activity of dislocations, denoted by ×, located inside the soft domains. (b) Stress distribution and relationships within the composite: in the soft domain of thickness $d_s$, $\tau_a = \tau_e + \tau_b$, where $\tau_e$ is the effective stress; in the hard domain of thickness $d_h$,

$\tau_a = \tau_n - \tau_f$, where $\tau_n$ is the net stress. A force balance across the hard/soft interface gives the relation $\tau_f d_h - \tau_b d_s = 0$, linking the internal stress components responsible for HDI strengthening.

However, the above practice raises two key questions that require clarification. First, what distinguishes the recently defined HS materials from previously known classes of heterogeneous materials, such as composites, dual-phase or multiphase alloys (e.g., steels, titanium alloys, and superalloys), functionally graded or gradient materials, precipitation-hardened alloys, polycrystals with bimodal or multimodal grain sizes, multilayers, or lamellar microstructures? A provisional answer could be that HS materials encompass and extend beyond these well-established categories [1]. This broad-stroke classification, nevertheless, could leave doubts as to whether homogeneous fine-grained or fine-twinned elemental metals could be counted as HS materials. This is because, when subjected to mechanical loading, the grains inside such a material are uniform in size but vary widely in crystallographic orientations, exhibiting very different Schmid factors on their respective maximally stressed slip systems. The pronounced grain-level variations in plastic response incur incompatibility across grain boundaries (GBs), giving rise to GNDs and, consequently, back stress, which is often regarded as a hallmark of HS materials [1]. A similar consideration applies to fine-twinned elemental metals. The second question, building on the first one, is whether a threshold degree of structural heterogeneity, or a minimum level of its effect on properties, is required for a material to qualify as HS. This remains an open question. If the observed property enhancements can already be explained by a rule-of-mixtures (ROM) average of the constituent zones making up the composite, introducing the HS as an additional label may be unnecessary. On the other hand, if heterostructuring introduces fundamentally new strengthening mechanisms, how significant is their contribution? Taking mechanical strength as an example, improvements are often anticipated in HS materials, consistent with established concepts such as back stress strengthening [9, 10] and kinematic hardening [11]. However, it remains unclear how to isolate and quantify the portion of strengthening specifically attributable to the designed (i.e., purposefully heightened) structural heterogeneity. This uncertainty stems from the fact that back stress can originate from multiple sources.

**Back stresses: Type I versus Type II**

In recent publications, most authors have evaluated $\sigma_b$ using Dickson's method [12], which analyzes the hysteresis loop from a tensile load-unload-reload (LUR) test to partition the total flow stress into back stress $\sigma_b$ and effective stress $\sigma_e$ components [2]. The measured $\sigma_b$ is typically attributed to HDI strengthening within HS materials. At first glance, this approach appears to provide a scalar metric for quantitatively assessing the effects of heterostructures. However, the sample-level $\sigma_b$ obtained from such measurements cannot be directly attributed to the intentionally introduced heterostructures, as $\sigma_b$ is a general feature that can arise from multiple sources. It is well established that during plastic deformation of otherwise homogeneous polycrystalline metallic materials [9], dislocation glide in different grains produces deformation incompatibility at GBs. If adjacent grains were to deform independently, such incompatibility would result in either overlap or opening. To accommodate this, GNDs accumulate at GBs [5]. Similar to the back stress in conventional dislocation cell structures [8], these GNDs associated with GBs produce long-range, directional back stresses that hinder further dislocation glide, giving rise to Hall-Petch-type hardening [13]. The GND density becomes particularly high when the grain size is refined, which is common in most recently developed HS materials (the grain size is often on nanoscale in the homogeneous hard A regions while much bigger in the much softer B regions). The ultrafine microstructure in these materials often result from multiple alloying elements introduced to promote copious nucleation while limiting the growth of recrystallized grains and phases [1-3]. As such, the high strength observed in HS materials largely originates from back stresses generated by plastically non-homogeneous deformation across neighboring grains (even when their sizes are uniform). In what follows, this will be referred to as Type I back stress, which is particularly high in ultrafine grained (UFG) metals, such as the 0.6-μm Al [14], or nanocrystalline (NC) metals, such as the 20-nm Ni [15], where $\sigma_b$ values exceeding 1 GPa have been reported. However, such a polycrystalline metal with variation in grain orientations and the presence of GBs is conventionally not considered a heterogeneous material from the microstructure standpoint.

In other words, although plastic deformation in a polycrystalline elemental metal is inherently heterogeneous, the material itself is not classified as an HS system. Moreover, its high strength, attributed to fine grain size, is already well described by classical GB strengthening mechanisms, such as dislocation pile-up giving the Hall-Petch relationship. In other words, Type I back stress and GB hardening reflect the same underlying mechanism in this case, both originating from GND accumulation at GBs.

However, there exists a second component in HDI strengthening, one that generates additional back stress beyond the Type I back stress already present in the constituent A and B regions. This extra component arises directly from the heterostructure itself, i.e., the intentionally introduced structural heterogeneity, as it happens only when the hard (A) and soft (B) regions are specially designed to generate a characteristic length scale $\lambda$. This $\lambda$ can represent the period of a gradually varying structure (e.g., sine wave) or the center-to-center distance between the A and B regions with a sharp interface. A structural gradient can then be defined, for example, as a hardness gradient of GPa/mm [16]. A high gradient corresponds to a large number of A/B couples per unit volume or length, leading to more (diffuse or sharp) A-B interfaces where plastic incompatibility must be accommodated. This, in turn, produces extra storage of additional GNDs, giving rise to what we define as the Type II back stress. As such, the total HDI stress measured via LUR testing comprises two distinct components: Type I and Type II back stresses. Crucially, only the Type II back stress reflects the "net value added" by heterostructuring—above the Type I contribution that is already inherent in homogeneously structured counterparts. This separation, which distinguishes the various microstructural sources contributing to back stresses, extends beyond earlier approaches that considered only an overall back stress for the material. The distinction between Type I and Type II back stresses, along with their respective magnitudes, can be more clearly illustrated in the following case study on nano-twinned (NT) Cu [16-18].

**Homogeneous versus heterogeneous nano-twinned Cu**

Four types of homogeneous NT (HNT) Cu samples were fabricated via direct-current electrodeposition, each featuring grains of uniform size and growth twins of uniform thickness increasing from 28, 37, 50 to 70 nm, referred to as HNT-A through HNT-D, respectively, as schematically illustrated in Fig. 2(a). Despite their structural homogeneity, LUR tests revealed distinct hysteresis loops for these samples (Fig. 2(b)), with the measured $\sigma_b$ increasing as the twin thickness decreases (Fig. 2(c)).

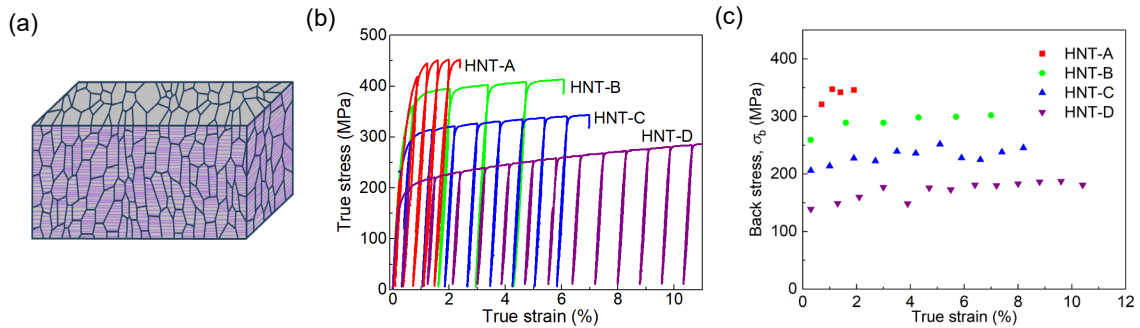

**Figure 2**. Microstructure, stress-strain behavior, and back/effective stress in HNT Cu, adapted from [16]. (a) Schematic of HNT Cu with uniform grain size and twin thickness. Each HNT Cu sample is characterized by growth twins of a specific thickness. HNT-A to HNT-D represent four different twin thicknesses. (b) Stress-strain curves of HNT-A to HNT-D with corresponding LUR hysteresis loops. For each HNT sample, the back stress $\sigma_b$ at various tensile strains is extracted from hysteresis loops measured in LUR tests. Notably, the unloading curve deviates markedly from the reference linear elastic unloading path. Particularly, reverse plastic yielding occurs while the applied stress remains tensile, indicating a strong Bauschinger effect associated with high back stress. These back stresses originate from GNDs accumulated at GBs/TBs in each HNT Cu sample. (c) The back stress $\sigma_b$ increases with decreasing twin thickness. These four types of HNT Cu will be combined in a stacked electrodeposition experiment to create a set of GNT Cu samples with prototypical heterogeneous nanostructures, as shown in Fig. 3.

Four types of gradient NT (GNT) Cu samples were also fabricated via direct-current electrodeposition by sequentially stacking the four homogeneous components, HNT-A through HNT-D. Each resulting GNT sample (GNT-1 to GNT-4, with GNT-3 schematically illustrated in Fig. 3(a)) contains an equal volume fraction (25%) of each component. However, the HNT components differ in slab thicknesses in GNT-1 to GNT-4, resulting in a different number of slabs per unit length along the growth direction (see detailed arrangements in [16]). This produces various degrees of structural difference per unit length, which we define as the structural gradient in this Viewpoint article.

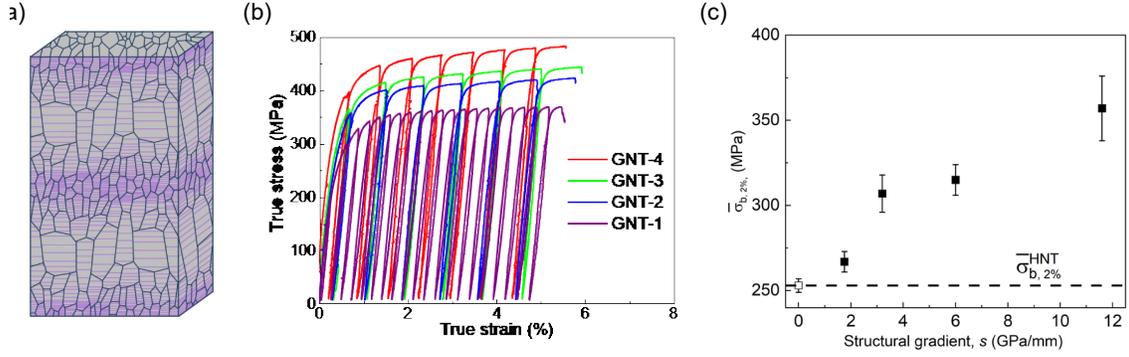

**Figure 3**. Microstructure, stress-strain behavior, and back/effective stress in GNT Cu, adapted from [16]. (a) Four types of HNT Cu, shown in Fig. 2, are sequentially deposited (see main text) to fabricate a set of GNT Cu samples with prototypical heterogeneous nanostructures. The schematic illustrates the structure of GNT-3, one of the four resulting GNT Cu samples (GNT-1 to GNT-4). (b) Stress-strain curves of GNT-1 to GNT-4 exhibit different strengths (corresponding to the sum of effective stress and back stress, with the latter being the dominant component) as well as LUR hysteresis loops. (c) The sample-level back stress $\bar{\sigma}_{b,2\%}$ at 2% strain, derived from the LUR loops, increases from GNT-1 to GNT-4 as the structural gradient (represented in terms of strength gradient) becomes more pronounced. The dashed line indicates the ROM estimate (253 MPa) for $\bar{\sigma}_{b,2\%}^{HNT}$, which is the weighted average of the $\sigma_b$ values from the four HNT Cu building blocks in Fig. 2(c). The ROM representation of the average Type I back stress is the same for all GNT-1 to GNT-4 samples [16].

Correspondingly, the sample-level back stress $\bar{\sigma}_b$, evaluated from the LUR hysteresis loops (Fig. 3(b)), increases with structural gradient. The four GNT Cu samples exhibit overall $\bar{\sigma}_{b,2\%}$ values (at 2% strain) ranging from 265 to 360 MPa. After subtracting the baseline back stress (253 MPa), which represents the total back stress in the limit of zero structural gradient, the excess $\bar{\sigma}_{b,2\%}$ ranges from 12 to 107 MPa. This extra back stress is defined as the Type II back stress, the origin of which is *bona fide* the intentionally introduced structural gradient, which increases from GNT-1 to GNT-4. In the case of GNT-4, which exhibits the highest $\bar{\sigma}_{b,2\%}$ (Fig. 3(c)), the Type II back stress accounts for 30% (107/360) of the total back stress. GNT-4 thus clearly qualifies as a heterogeneous material, as its pronounced spatial structural gradient and the corresponding strain gradient give rise to a significant Type II back stress. In fact, the excess strength of GNT-4 beyond the ROM prediction primarily arises from the Type II back stress [16]. In contrast, at a structural gradient of ~2 GPa/mm, the material is not all that heterogeneous after all, as the contribution from the Type II back stress is only 4.5% (12/265). This case study underscores the importance of decomposing the overall back stress into Type I and Type II components, providing a quantitative measure of the "value added" by heterostructuring.

**Defining structural gradient hardening**

Building on the preceding analysis, we define *structural gradient hardening* (SGH) as the extra strengthening represented by the Type II back stress, beyond the Type I back stress inherent to the homogeneous constituents assembled in an HS material. SGH therefore quantifies the strengthening that uniquely arises from a characteristic structural (strain or hardness) gradient introduced through intentional heterostructuring, a component not captured by the ROM or conventional strengthening mechanisms. An example is shown in Fig. 3(c), where the SGH contribution is determined by subtracting the ROM-estimated

Type I stress (dashed line) from the experimentally measured total back stress (y-axis). The former is obtained as the weighted average of the Type I stresses of the individual A and B constituents, experimentally measured from uniformly structured A or B samples sharing the same composition and microstructure as the A or B regions in the HS material.

Three factors justify the introduction of this new concept of SGH, and its utility is clearly demonstrated in the case study of GNT Cu discussed above. First, SGH highlights enhancements in properties that go beyond predictions from the traditional ROM approach, which is typically applied to known classes of heterogeneous materials such as composites. By requiring a distinct extra hardening contribution, the SGH definition excludes materials that lack pronounced structural heterogeneity, in particular single-phase (or single-element) metals with uniform or narrowly distributed microstructure (e.g., grain or twin) sizes, or coupled A/B parts with few interfaces. In this context, GNT Cu embodies SGH, while HNT Cu does not. Second, the SGH framework requires a quantitative assessment of how much the heterostructure contributes to property enhancement. Specifically, SGH, as captured by the Type II back stress, isolates the extra gain beyond the Type I back stress expected from a homogeneous distribution of strengthening features (e.g., TBs or GBs, such as in uniformly structured HNT Cu in Fig. 2). In contrast, reporting only the total strength (and total back stress), as is often the case in recent HS literature, fails to distinguish the strengthening due to heterostructuring from that due to conventional mechanisms. Third, the separation of Type I and Type II back stress components provides a direct link between mechanical properties and their structural origins, especially for Type II, which directly reflects the effects of intentional heterostructuring. This causal and quantitative structure-property relationship, often lacking in current discussions of HS materials, is critically needed to advance the field from a materials science standpoint.

**Analysis via representative volume elements**

Let us now recapitulate the preceding concepts and examine the effect of SGH using a multiscale approach. At this point, we move beyond sample-level descriptions to consider both structural and stress distributions. Figure 4 presents a schematic to illustrate our multiscale approach, examining both large and small representative volume elements (RVEs) at different length scales [16-18]. For the heterostructured NT Cu as a model system, the entire gradient-structured sample of GNT Cu is treated as a "large" RVE. In contrast, the zoomed-in blue box in Fig. 4 shows a "small" RVE representing a "soft" region with uniformly thick twin lamellae, adjacent to another small RVE representing a "hard" region with uniformly thin twin lamellae. The back stress at the large RVE level represents an average of the back stresses of the constituent small RVEs. Each small RVE is, first of all, strengthened by a Type I back stress that arises from the TBs within. Although each small RVE is structurally homogeneous in terms of twin thickness, it still undergoes non-homogeneous plastic deformation due to orientation-dependent variations in Schmid factors across the twin lamellae. During plastic loading, this deformation incompatibility at TBs is accommodated by the accumulations of GNDs. In such nanostructured cases, the high strength is largely attributed to Type I back stress, which originates from incompatible deformation between neighboring nanoscale lamellae. Importantly, the strengthening effect increases with decreasing twin thickness as expected from Hall-Petch-type hardening, whereas the effective stress component remains relatively insensitive to lamella thickness [19].

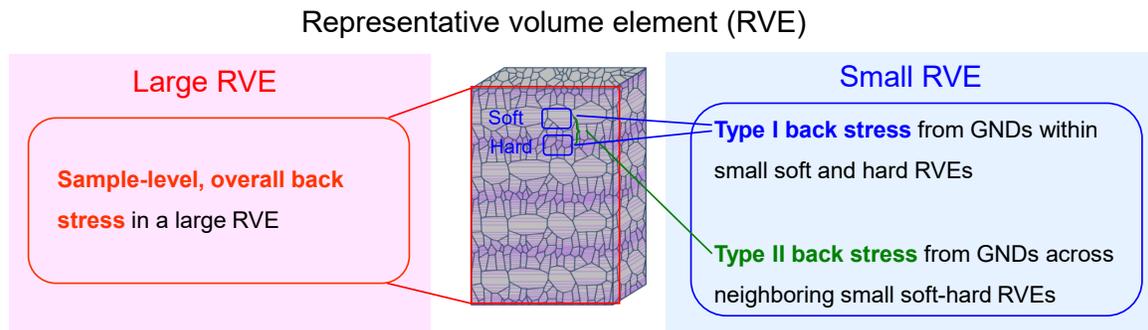

**Figure 4**. The distinct roles of large and small RVEs are highlighted in the analysis of back stress strengthening in GNT Cu, adapted from [16-18]. Tensile loading is applied along the horizontal direction.

Meanwhile, the Type II back stress within each small RVE arises from structural gradients between adjacent small RVEs (i.e., HNT Cu), when these building blocks are stacked into a GNT Cu configuration with a vertical gradient in twin thickness. Each small RVE possesses a yield strength determined by its internal homogeneous twin thickness; differences in these strengths create a spatial gradient of plastic strain in GNT Cu under loading. This gradient, in turn, induces additional GNDs to accommodate the resulting deformation incompatibility. This situation is analogous to bending in a beam, where the top and bottom surface regions experience larger plastic strains than the neutral-plane region in the middle of the beam, requiring extra GNDs to accommodate the lattice curvature associated with gradient plastic strains. In GNT Cu, these GNDs add long-range, directional back stresses that hinder dislocation glide within each small RVE. The cumulative effect is an additional sample-level back stress, i.e., Type II back stress, arising from heterostructuring in GNT Cu. This extra strengthening effect is quantitatively captured by the nonlocal theory of strain gradient plasticity [16-18]. While the theory was originally developed to explain strain gradient hardening induced by externally imposed non-uniform deformation, such as torsion, bending, and indentation [20, 21], on otherwise homogeneous microstructures, it now extends to materials characterized by heterogeneous microstructures that introduce additional cross-RVE-level plastic heterogeneity [16-18]. The same framework can be extended to polycrystals with bimodal or multimodal grain size distributions that promote extra GNDs [22], where GBs play a role similar to TBs in GNT Cu. In such cases, grain size replaces twin thickness as the structural parameter distinguishing the hard and soft domains.

**Adding up strengthening terms**

The analysis above shows that the principal effect of heterostructuring is to generate extra strength by enhancing plastic strain gradient, which lead to the accumulation of additional GNDs and the resultant Type II back stress. Distinguishing Type II-contributed SGH from Type I back stress is crucial, as confusion and misrepresentation frequently appear in recent literature. In particular, some studies mistakenly sum multiple strengthening contributions without recognizing overlap. For instance, the HDI strengthening term from the total stress, obtained from LUR measurements, already includes contributions from GNDs responsible for Type I back stress, which frequently dominates the overall back stress. This point can be illustrated by returning to heterostructured Cu as an example. This HS material was designed by integrating several homogeneous HNT Cu building blocks, each initially lacking a heterostructure. The combination of those building blocks yields, for example, GNT-4 [16-18], whose post-yield strength is primarily governed by a total HDI stress of 360 MPa. Of this, 253 MPa originates from Type I back stress due to reduced twin thickness (Hall-Petch-type hardening), while the remaining 107 MPa is attributed to the Type II back stress introduced by heterostructuring. With this breakdown, it is evident that caution must be exercised to avoid double-counting HDI strengthening and GB/TB strengthening. Unfortunately, in recent literature, these two are sometimes cited as independent, additive contributions. In the example above, it would clearly be incorrect to claim that GNT-4 benefits from "360 MPa back stress plus TB/GB hardening", since the latter is already embedded in the former.

As a thought experiment to further clarify the difference between GB/TB hardening and SGH, consider if the GNT Cu had been designed to consist almost entirely of a single HNT variant, e.g., HNT-A, with a volume fraction approaching 100% instead of ~25%. Then the total back stress would be dominated by the Type I back stress in HNT-A (~350 MPa, Fig. 2(c)), primarily arising from Hall-Petch-type hardening associated with the ultrathin twin thickness (28 nm). Although this value is comparable in magnitude to the 360 MPa of GNT-4, it would lack a significant contribution from Type II back stress. As such, this material would not qualify as an SGH system.

**Concluding remarks**

In summary, inspired by ongoing research on HS materials, our analysis in this Viewpoint article identifies a niche but critical contribution, namely, SGH. Both key questions raised at the beginning of this article can then be satisfactorily addressed from this standpoint. First, for many HS materials, their properties are below, or well described by, the ROM average of the constituents. In such cases, the performance does not truly stem from intelligent heterostructuring, and existing ROM estimates suffice, making the additional HS label inconsequential.

We advocate for the quantitative evaluation of Type II back stress. This is the component that arises specifically from intelligent heterostructuring, distinct from Type I back stress, which originates from the distribution of conventional strengthening features, such as GBs and interfaces, within otherwise homogeneous regions of HS materials. The practice in recent publications of measuring only the total back stress is insufficient to rigorously justify the claimed benefits of heterostructures over their homogeneous counterparts. Separating Type II from Type I is essential for establishing a direct causal relationship between the engineered structural gradient and the SGH it induces. It is the structural gradient that necessitates the accumulation of additional GNDs to accommodate the associated plastic strain gradients, thereby generating the extra (Type II) back stress, which in turn contributes excess strength. Note that this contribution is not accounted for by the ROM, which has been adequate for many previously reported heterogeneous materials. Moreover, the Type II back stress also improves strain hardening and ductility, stemming from the continuous storage of GNDs due to structural gradients and the corresponding increase in Type II back stress with increasing applied load. The SGH mechanism defined above should be the telltale trait that makes HS materials uniquely compelling, distinct from their homogeneously structured counterparts and from conventional composites or coarse-level heterogeneous materials that rely primarily on textbook strengthening mechanisms.

Looking ahead, we call for in-depth research into materials with multi-level structures, where multiple sources of back stresses arise from structural heterogeneities across distinct characteristic length scales [16, 23]. Particular attention should be given to the design of novel heterostructures that simultaneously achieve both high structural gradients and refined microstructures. The former refers to reduced length scales between soft and hard regions, which amplify Type II back stress (i.e., SGH), while the latter involves decreasing microstructural feature sizes within both soft and hard regions, thereby enhancing Hall-Petch-type hardening and contributing to increased Type I back stress. Nanostructures engineered with high structural gradients are especially promising for achieving this dual objective [6], as they promote both strengthening mechanisms in tandem.

As a general guideline for designing HS materials, the first step is to select the "raw materials" for the hard (A) and soft (B) zones, including factors that control hardness or softness (e.g., grain size) and their overall volume fractions (e.g., 60%-40%). The expected performance can be estimated using known mechanisms and empirical formulas such as the ROM. This step of materials selection establishes the baseline (Type I) back stress, serving as the foundation for Step 2, which is the more intellectually demanding process of heterostructuring. By tailoring the size, number, shape, and spatial distribution of the A and B zones, the structural gradient can be optimized to maximize Type II back stress, thereby enhancing mechanical performance. We expect that the insights presented in this Viewpoint article will inspire future studies to more fully exploit the opportunities provided by SGH and help identify the "low-hanging fruit" in designing HS materials with superior mechanical performance.